# SurroPilot: An LLM-Assisted Platform for Heterogeneous Surrogate Endpoint Evaluation in Clinical Trials

Xingyu Li[†,*][1] and Peng Wei[†][1]

[1]Department of Biostatistics, The University of Texas MD Anderson Cancer Center, 77030, Houston, TX, U.S.A.

[*] Corresponding author: xli36@mdanderson.org
[†] These authors contributed equally to this work.

## Abstract

Surrogate endpoints are widely used in clinical trials to accelerate treatment evaluation, yet their validity may vary substantially across patient subgroups. Although recent advances in heterogeneous causal mediation analysis enable subgroup-specific surrogate evaluation, applying these methods requires substantial expertise in causal inference, statistical programming, and clinical trial methodology, limiting their accessibility to many biomedical researchers. We present SurroPilot, a large language model (LLM)-assisted platform for heterogeneous surrogate endpoint evaluation in clinical trials. Through natural-language interaction, SurroPilot supports the complete analytical workflow, including dataset understanding, data preprocessing, mediator and covariate selection, heterogeneous causal mediation analysis, subgroup interpretation, and automated report generation. To improve the reliability of AI-assisted statistical computing, the platform incorporates a shared-context programmer–inspector framework for iterative R code correction and automated validation of LLM-generated variable selections. Rather than replacing statistical methodology, SurroPilot integrates LLM with a validated heterogeneous mediation framework, allowing the LLM to assist with analytical reasoning while statistical inference is performed using established causal inference methods. Using the ACTG175 Phase III HIV clinical trial, we demonstrate that SurroPilot provides an end-to-end, reproducible workflow for heterogeneous surrogate endpoint evaluation and substantially lowers the technical barriers to applying advanced causal mediation methods in clinical trial research.

Keywords Heterogeneity, machine learning, large language models, surrogate biomarker, survival analysis

## 1. Introduction

Evaluating the clinical benefit of a new therapeutic intervention, defined by improvements in survival, symptoms, or functional outcomes, often requires prolonged follow-up, particularly in chronic or lifethreatening diseases. To address the inherent delays associated with measuring such endpoints, the U.S. Food and Drug Administration (FDA) established the Accelerated Approval pathway in 1992 [FDA, 1992]. This regulatory mechanism permits approval of drugs for serious conditions with unmet medical needs on the basis of surrogate or intermediate clinical endpoints, thereby expediting patient access to potentially effective therapies. A surrogate endpoint is a biomarker, such as a laboratory measurement, radiographic finding, or physical sign, that is intended to predict clinical benefit but does not itself directly measure it [FDA, 2014]. An intermediate clinical endpoint similarly reflects a therapeutic effect that is reasonably likely to forecast ultimate clinical benefit, for example through its influence on irreversible morbidity or mortality. By substituting such endpoints for definitive outcomes, the accelerated approval framework can substantially shorten the drug development timeline. In oncology, for instance, regulatory approval may be granted based on tumor response rather than overall survival (OS), under the assumption that tumor shrinkage is reasonably likely to predict meaningful clinical benefit.

Despite the widespread use of surrogate endpoints in drug development, existing evaluation frameworks generally treat surrogate validity as a homogeneous, population-level property. This assumption is rarely examined and may not hold in the presence of substantial biological and clinical heterogeneity across patients. Consequently, a surrogate biomarker that reliably predicts clinical benefit for one subgroup may perform poorly for another, limiting the generalizability of population-level validation. This limitation has important regulatory implications because accelerated approval based on surrogate endpoints requires subsequent confirmatory trials to verify that improvements in the surrogate translate into meaningful clinical benefit, such as OS. However, an increasing number of confirmatory studies have failed to demonstrate the anticipated clinical benefit. To date, FDA has withdrawn accelerated approval for 31 oncology drugs, including 25 since 2020, as well as nine additional drugs in non-oncology indications. In response, between 2023 and 2025 the FDA issued a series of draft guidances strengthening the evidentiary requirements for accelerated approvals based on surrogate endpoints, including the expectation that confirmatory trials be underway at the time of approval [FDA, 2023, 2024, 2025]. These developments underscore the need to move beyond population-level assessments and evaluate surrogate validity at the patient subgroup level.

Causal mediation analysis has emerged as a principled statistical framework for evaluating surrogate biomarkers by quantifying the extent to which treatment effects are transmitted through intermediate endpoints [Vandenberghe et al., 2018, Zhou et al., 2021, 2022, 2023]. Unlike traditional correlationbased approaches, causal mediation analysis enables a mechanistic assessment of surrogate validity by decomposing treatment effects into direct and indirect components under explicit causal assumptions. Building upon this framework, recent methodological advances have extended mediation analysis to account for patient heterogeneity. In particular, Li et al. [2026] proposed a heterogeneous survival mediation framework that identifies clinically interpretable patient subgroups with differential surrogate validity, allowing surrogate biomarkers to be evaluated at the subgroup level rather than assuming a single population-wide relationship. These developments provide a more nuanced and clinically meaningful framework for surrogate endpoint evaluation. Collectively, these advances provide a more flexible and clinically meaningful framework for surrogate endpoint evaluation and have the potential to improve precision medicine by identifying patient populations that benefit through distinct biological pathways.

Despite these methodological advances, applying heterogeneous causal mediation analysis in clinical trials remains challenging in practice. A rigorous analysis requires a sequence of complex statistical decisions, including identifying biologically plausible surrogate biomarkers, selecting appropriate covariates, specifying causal models that satisfy underlying identification assumptions, and determining clinically meaningful patient subgroups. In addition, researchers must perform extensive data preprocessing, implement specialized statistical methods, assess model diagnostics, and interpret often complex subgroup-specific mediation results. These tasks demand substantial expertise in causal inference, biostatistics, and clinical trial methodology, creating a steep learning curve for investigators without specialized training. Consequently, advanced surrogate evaluation remains largely inaccessible to many biomedical and clinical researchers, limiting the broader adoption of heterogeneous mediation methodologies in practice.

Recent advances in artificial intelligence (AI), particularly LLMs, offer a promising opportunity to lower the technical barriers associated with advanced statistical analyses by enabling naturallanguage interaction with complex analytical workflows. Regulatory agencies, including the FDA, have increasingly recognized the potential of AI and machine learning to support clinical trial design, conduct, data management, endpoint assessment, and clinical trial analytics [FDA, 2025]. Existing AI-assisted applications have demonstrated considerable value in improving research efficiency through tasks such as protocol development, patient recruitment, eligibility screening, data extraction, and automated documentation [Olawade et al., 2026]. However, these systems have largely focused on operational and administrative workflows, while the use of AI to support statistical inference, causal reasoning, and interpretable clinical trial analyses remains limited. As a result, AI has yet to realize its full potential in enabling rigorous and accessible statistical analyses for modern clinical research.

To address these challenges, we propose SurroPilot, an AI-assisted statistical framework for heterogeneous surrogate endpoint evaluation that integrates LLMs with validated causal mediation methodology. Rather than replacing statistical methodology or human expertise, SurroPilot adopts a human-in-the-loop paradigm in which LLMs facilitate data understanding, variable selection, statistical computation, and interpretation, while all causal inference is performed using validated statistical procedures. The main contributions of this work are summarized as follows. First, we present, to the best of our knowledge, the first AI-assisted framework for heterogeneous surrogate endpoint evaluation in clinical trials. By combining natural-language interaction with heterogeneous causal mediation analysis, SurroPilot substantially lowers the technical barrier to applying advanced causal inference methods while preserving statistical rigor and reproducibility. Second, we develop an LLM-assisted statistical workflow that automates the major analytical components of heterogeneous mediation analysis, including data understanding, preprocessing, mediator and covariate selection, heterogeneous subgroup identification, result interpretation, and report generation. This unified workflow enables reproducible end-to-end analyses from natural-language instructions rather than manually written statistical code. Third, we introduce a shared-context programmer–inspector self-correction mechanism that iteratively refines LLM-generated R code and validates variable selections through execution feedback. By maintaining a shared conversational context throughout the analytical process, the proposed framework improves the reliability and robustness of AI-assisted statistical computing. Finally, we evaluate SurroPilot using the ACTG175 Phase III clinical trial and demonstrate that state-of-the-art LLMs can reliably recover expert-defined mediators and adjustment covariates and reproduce heterogeneous subgroup analyses consistent with expert-guided analyses. These results demonstrate the feasibility of integrating large language models with modern causal inference methodology for practical clinical trial analysis.

## 2. Background and Related Works

### 2.1. Surrogate Biomarkers and Heterogeneous Surrogacy

Surrogate endpoints are post-treatment variables intended to capture the effect of an intervention on a clinically meaningful outcome, particularly in settings where definitive endpoints such as OS are rare, delayed, or costly to observe. By leveraging treatment effects on surrogate endpoints, investigators aim to infer treatment effects on the true clinical outcome, thereby accelerating treatment evaluation and regulatory decision making [Buyse et al., 2010]. Consequently, surrogate endpoints have become increasingly important in modern clinical trials and drug development programs. Examples include pathological complete response in breast cancer and objective response rate (ORR) in oncology, both of which have been used to support regulatory approval and treatment evaluation [FDA, 2025].

A fundamental challenge in surrogate evaluation is that surrogate validity may not be homogeneous across patient populations. Biological heterogeneity, treatment-response mechanisms, and patientspecific characteristics can all influence the extent to which treatment effects on a surrogate endpoint translate into effects on the true clinical outcome. Evidence from oncology studies has demonstrated that surrogate relationships may vary substantially across molecularly defined subgroups, indicating that a biomarker considered informative at the population level may exhibit markedly different predictive performance in specific patient subsets [Cohen et al., 2003, Kazandjian et al., 2016]. These observations have motivated increasing interest in heterogeneous surrogacy, where the goal is to identify subgroups for which a candidate biomarker serves as a reliable surrogate and distinguish them from subgroups in which the surrogate relationship is weak or absent. In Figure 1, (a) represents the ideal surrogate scenario, where the surrogate can transmit all treatment effects, (b) represents the surrogate can only reflect part treatment effects, which are the most of the scenarios, and (c) represents the surrogate can not reflect the treatment effects, in this scenario; it is not true surrogate. However, these scenarios can emerge in the same trial, leading to surrogate heterogeneity.

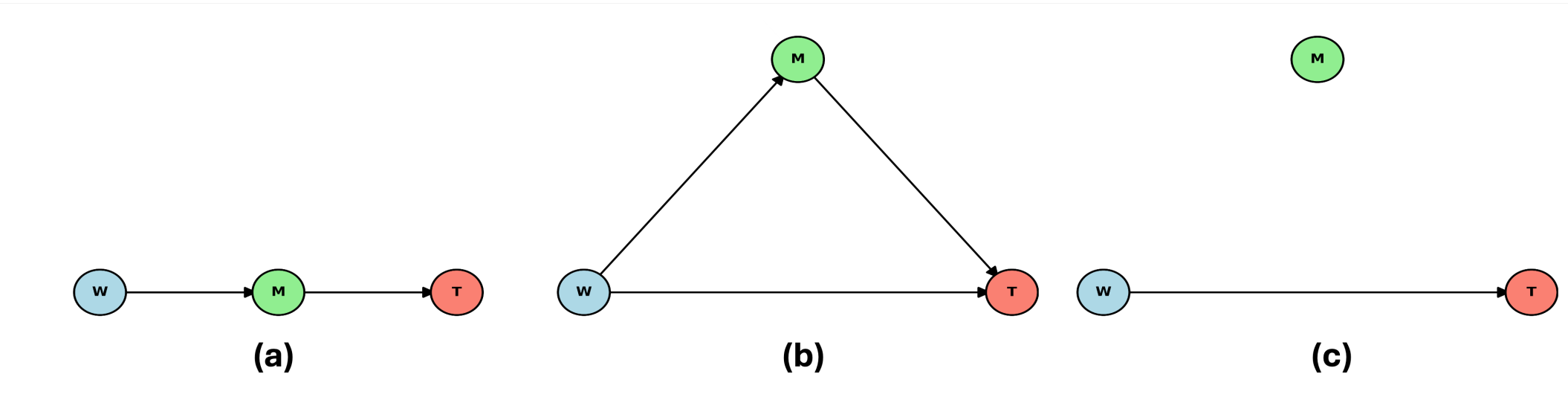

Figure 1 Causal diagrams for surrogate endpoint validation and mediation model. Panels (a)–(c) depict three scenarios of surrogate relationships. These scenarios provide illustrative examples for the analytical workflow presented in panel (d). (a) represents an ideal surrogate scenario, in which the surrogate endpoint $M$ fully captures the treatment effect on the clinical outcome $T$. (b) represents a non-ideal surrogate scenario, in which the surrogate endpoint captures only part of the treatment effect, with the remaining effect acting directly on the clinical outcome. $W$ denotes treatment, $M$ denotes surrogate biomarker, $T$ denotes clinical outcome. (c) represents $M$ is not surrogate endpoint, with no treatment effect mediated by $M$.

## 2.2. Causal Mediation Analysis for Surrogate Evaluation

Causal mediation analysis provides a principled framework for evaluating surrogate biomarkers by quantifying the extent to which treatment effects are transmitted through intermediate endpoints. The methodology has been extensively developed for survival outcomes. VanderWeele [2011] established a mediation framework under the Cox proportional hazards model and defined direct and indirect effects on the hazard scale. Subsequently, Huang and Yang [2017] extended survival mediation analysis to accommodate multiple mediators, while Zhang et al. [2021] generalized the framework to highdimensional settings. More recently, Chi et al. [2024] proposed an $R^2$-based measure for quantifying mediation effects in survival analysis.

Although these methods provide powerful tools for surrogate evaluation, most existing approaches focus on population-level mediation effects and implicitly assume homogeneous surrogate validity across individuals. To address this limitation, Li et al. [2026] proposed a heterogeneous survival mediation analysis framework that identifies patient subgroups exhibiting differential mediation effects and subgroup-specific surrogate performance. By characterizing heterogeneity in surrogate validity, the framework enables a more nuanced understanding of treatment mechanisms and supports precision-medicine applications.

Despite these methodological advances, implementing heterogeneous mediation analysis remains challenging in practice. Successful application requires careful mediator and covariates selection, appropriate model specification, subgroup identification, diagnostic assessment, and interpretation of complex causal estimands. These tasks often depend on substantial statistical expertise and domain knowledge, and typically involve extensive programming and iterative model refinement. Consequently, advanced mediation methods remain difficult to deploy routinely in clinical trial research.

## 2.3. AI-assisted Clinical Trial Analytics

Recent advances in AI, particularly LLMs, have created new opportunities to support clinical trial design, conduct, and analysis. Regulatory agencies have increasingly recognized the potential of AI to facilitate disease characterization, identify clinically meaningful patient subgroups, improve endpoint assessment, and enhance clinical trial analytics. Recent FDA guidance further emphasizes that AI-generated evidence can contribute to regulatory decision-making when accompanied by appropriate credibility assessment and human oversight [FDA, 2025].

The emergence of LLMs has substantially expanded the capabilities of AI-assisted systems beyond traditional task automation. Through their abilities in natural-language understanding, knowledge retrieval, reasoning, and code generation, LLM-powered agents are increasingly capable of interacting with complex analytical workflows and assisting users in solving domain-specific problems. Consequently, LLM-based assistants have been successfully applied in diverse areas, including software engineering, scientific discovery, healthcare, and biomedical research [Wu et al., 2023, Team et al., 2023, Yang et al., 2025].

Within clinical research, existing AI-assisted systems have primarily focused on operational and administrative tasks, such as protocol development, patient recruitment, eligibility screening, clinical documentation, literature synthesis, and clinical decision support. These applications have demonstrated the potential of LLMs to improve research efficiency and reduce the burden associated with increasingly complex clinical studies. However, the application of LLMs to statistical inference and causal analysis remains largely unexplored. In particular, no existing AI-assisted framework is specifically designed to support heterogeneous causal mediation analysis and surrogate endpoint evaluation, despite the substantial methodological complexity and practical importance of these analyses in modern clinical trials.

## 2.4. Research Gap and Positioning

To address these challenges, we propose an AI-assisted framework "SurroPilot" for heterogeneous surrogate biomarker evaluation in clinical trials. The framework integrates an LLM-based analytical assistant with a validated heterogeneous mediation analysis pipeline, enabling users to perform complex causal mediation analyses through natural-language interaction. Rather than replacing statistical methodology or human judgment, the proposed system adopts a human-in-the-loop paradigm in which LLMs assist with mediator selection, covariates identification, model configuration, interpretation of subgroup-specific mediation effects, and automated report generation.

By combining AI-driven analytical guidance with established causal inference methods, the framework seeks to reduce the technical barriers associated with heterogeneous surrogate evaluation while maintaining transparency, reproducibility, and methodological rigor. More broadly, this work extends the role of LLMs in clinical trials from operational support toward AI-assisted statistical discovery, providing a practical pathway for translating advanced causal mediation methodologies into routine clinical research practice.

# 3. Methods

Our proposed LLM-assisted platform, SurroPilot, provides an end-to-end workflow for heterogeneous surrogate endpoint evaluation in clinical trials through natural language interaction. Rather than requiring users to manually preprocess data, specify statistical models, or write analysis code, SurroPilot orchestrates a sequence of AI-assisted analytical modules that transform high-level user instructions into a complete and reproducible mediation analysis pipeline.

As illustrated in Figure 2, the framework consists of six sequential stages. First, users describe the clinical trial and analysis objectives in natural language. The LLM then interprets the dataset, identifies variable types, generates executable R code for data cleaning, and iteratively corrects execution errors until an analysis-ready dataset is obtained. Next, the LLM assists in selecting candidate surrogate biomarkers (mediators) and clinically relevant covariates by integrating user-provided descriptions with domain knowledge. The selected variables are subsequently passed to the heterogeneous causal mediation analysis module, where the heterogeneous causal mediation analysis method is executed to estimate subgroup-specific indirect treatment effects and identify clinically interpretable patient subgroups. Finally, the platform automatically summarizes the analytical findings, provides statistical interpretation, and generates a structured report containing selected variables, subgroup characteristics, estimated mediation effects, and clinical insights.

Unlike conventional analysis pipelines that require substantial programming expertise, SurroPilot decouples domain knowledge acquisition from statistical computation by tightly integrating human expertise with LLM-assisted reasoning. Domain knowledge is not solely inferred by the LLM; instead, it can be explicitly specified by domain experts and iteratively refined through human–AI collaboration. The LLM functions as an intelligent interface for data understanding, reasoning, variable selection, code generation, and result interpretation, whereas the underlying statistical engine performs heterogeneous causal mediation analysis using validated statistical methodology. This collaborative design enables clinicians and biomedical researchers to perform complex surrogate endpoint analyses through natural language while preserving methodological rigor, reproducibility, and expert oversight.

## 3.1. User Inputs

After uploading the dataset, users provide a natural language description of the data. Rather than specifying statistical models or analysis parameters, they describe the semantic meaning of variables, the study background, and other relevant domain knowledge that may inform downstream analyses. This natural language interface enables researchers to interact with the system intuitively, without requiring programming expertise or predefined analytical syntax.

To facilitate accurate interpretation of the dataset, SurroPilot automatically extracts structural metadata from the uploaded data and incorporates them, together with the user-provided description, into a task-specific system prompt. The extracted metadata include the dataset dimensions, variable names, data types, missing-value summaries, and descriptive statistics for each variable. By combining semantic descriptions with statistical metadata, the prompt provides the LLM with comprehensive contextual information for understanding the dataset before downstream analysis.

The resulting prompt establishes the execution context, specifies the expected input and output formats, and provides representative examples for in-context learning. This structured prompt enables the LLM to interpret the dataset consistently and generate structured outputs tailored to the subsequent data understanding and preparation workflow. The complete system prompt used in this stage is illustrated in Appendix Figure 1.

## 3.2. AI Data Understanding & Preparation

Following the user input stage, SurroPilot performs AI-assisted data understanding and preparation. Based on the user-provided dataset description and the extracted dataset metadata, the LLM first interprets the dataset structure and identifies the preprocessing operations required for downstream analysis. Typical operations include filtering observations according to user-defined criteria, handling missing values, recoding variables, transforming data types, and constructing derived variables when appropriate.

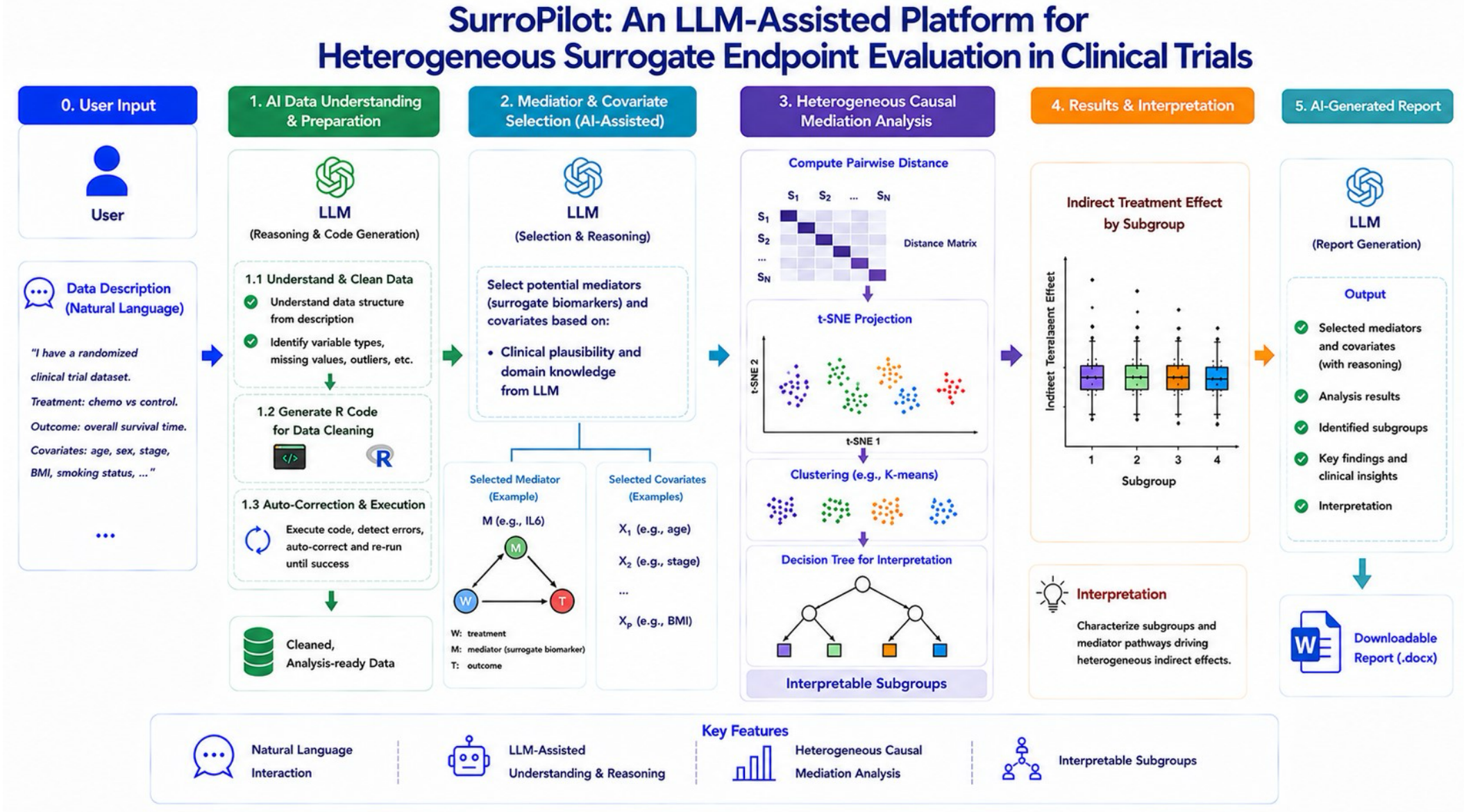

Figure 2 Overview of the SurroPilot framework. SurroPilot is an LLM-assisted platform for heterogeneous surrogate endpoint evaluation in clinical trials. Users interact with the system using natural language to describe the clinical trial dataset. The LLM first understands the dataset and automatically generates R code for data cleaning with iterative error correction. It then identifies candidate surrogate mediators and adjustment covariates based on clinical plausibility and domain knowledge, while providing transparent reasoning for each selection. The cleaned data are subsequently analyzed using the proposed heterogeneous causal mediation analysis framework, which computes pairwise distances based on individual indirect treatment effects, projects patients into a lowdimensional representation to reveal subgroup structure, and identifies clinically interpretable subgroups. Finally, SurroPilot summarizes the selected variables, analytical results, subgroup characteristics, and interpretations into an automatically generated report.

After identifying the required preprocessing operations, the LLM assumes the role of a programmer and generates executable R code to implement the requested data preparation tasks. The generated code is executed automatically within the R environment, enabling an end-to-end data preparation workflow without requiring manual programming. Because LLM-generated code may occasionally contain syntactic or logical errors, SurroPilot incorporates an iterative self-correction mechanism to improve execution reliability and robustness.

Specifically, if code execution fails, the LLM switches from the programmer role to an inspector role. The inspector analyzes both the generated R code and the corresponding execution error messages to identify the underlying causes of failure and produces targeted revision suggestions. These suggestions are then provided to the programmer, which regenerates corrected code for re-execution. This programmer–inspector cycle is repeated until the code executes successfully or a predefined maximum number of iterations is reached.

Unlike conventional multi-agent frameworks that employ separate LLM instances, the programmer and inspector in SurroPilot operate within a shared conversational context. As a result, the complete interaction history, including the user instructions, generated code, execution results, and revision feedback, is retained throughout the iterative process. This shared context facilitates more accurate error diagnosis, reduces repeated failures, and improves the robustness and consistency of automated data preparation. The workflow of this programmer–inspector framework is illustrated in Figure 3, and an example of the corresponding self-correction prompt is provided in Appendix Figure 2.

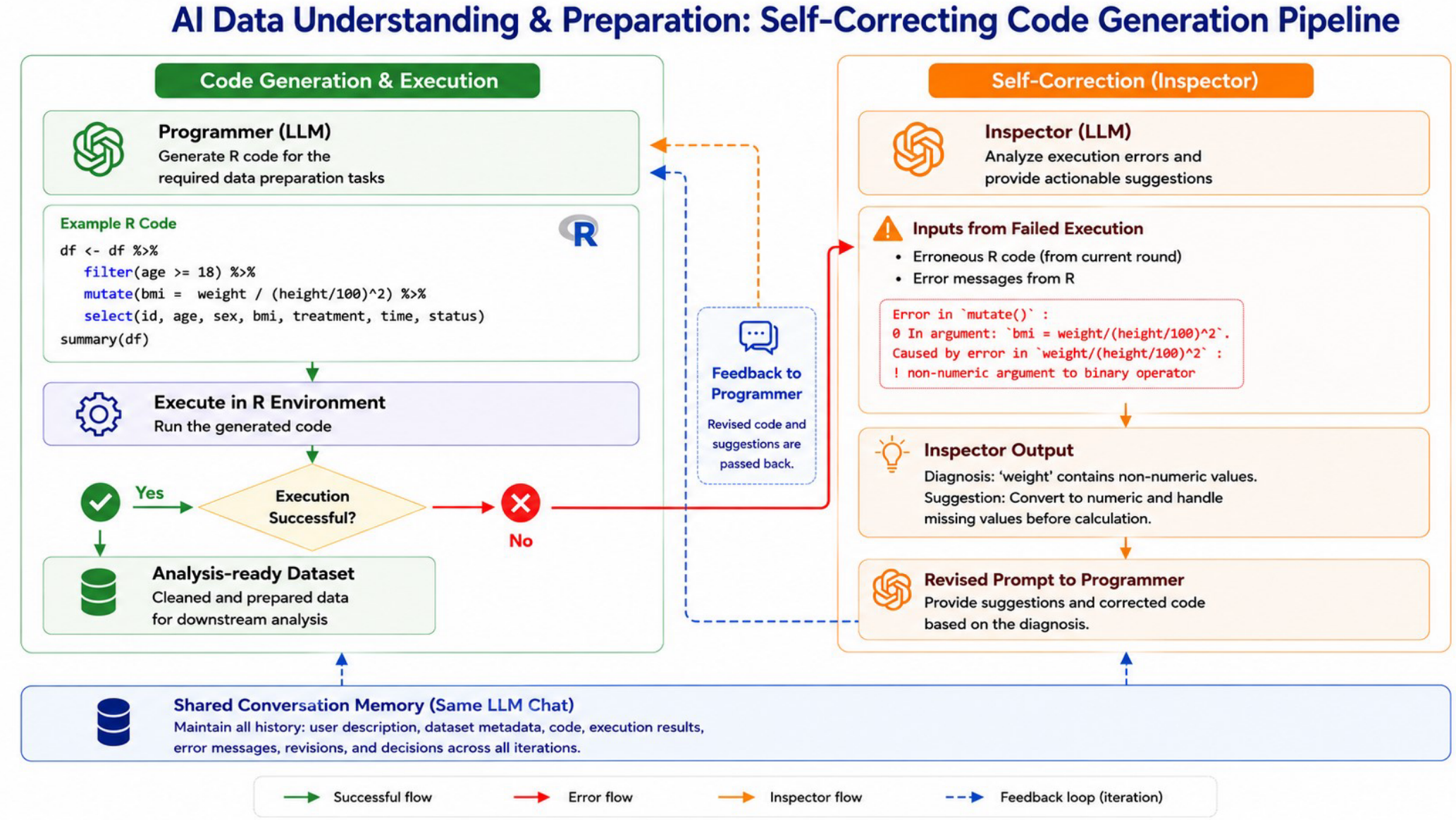

Figure 3 Following user instructions, the programmer role of the LLM generates executable R code for data preparation, which is automatically executed in the R environment. If execution fails, the inspector role analyzes the generated code together with the corresponding error messages and provides targeted revision suggestions. These suggestions are fed back to the programmer for code regeneration, forming an iterative feedback loop until successful execution or a predefined maximum number of attempts is reached. Both roles operate within the same conversational context, enabling shared memory of user instructions, generated code, execution results, and revision history throughout the entire workflow.

## 3.3. Mediator and Covariate Selection

Following data preparation, the users specifie the treatment variable, outcome variable (including the survival time and censoring status variables), and the corresponding treatment, control, and censoring levels. Before proceeding, the system automatically verifies that all user-specified variables are present in the uploaded dataset and that the specified levels are valid, thereby preventing inconsistencies between user input and the underlying data.

To incorporate domain expertise into the analytical workflow, users may optionally provide additional clinical or biological information, such as prior knowledge of disease mechanisms, hypothesized mediating pathways, or variables known to be associated with the treatment or outcome. This information, together with the dataset description and variable metadata, is provided to the LLM, which recommends candidate mediators and adjustment covariates. For selected variables, the LLM also generates a brief rationale based on causal mediation principles and the user-provided contextual information, enabling transparent documentation of the variable selection process. The example of prompt can be found at Appendix Figure 3.

Because LLM-generated outputs may occasionally contain variables that are absent from the dataset, use alternative variable names, or produce selections that are inconsistent with the available data, the proposed programmer–inspector framework performs an automated verification step. The inspector compares every selected mediator and covariate against the uploaded dataset, validates variable names and data availability, and identifies inconsistencies or unsupported selections. When an error is detected, corrective feedback is returned to the programmer, which revises the selection while preserving the complete conversational history. This iterative process continues until all selected variables are successfully validated. The example of prompt can be found in Appendix Figure 4.

The final output of this stage consists of a validated set of treatment, outcome, mediator, and adjustment covariates, together with the corresponding selection rationale. These validated inputs are subsequently used by the heterogeneous causal mediation analysis framework, ensuring that downstream analyses are performed on a consistent and reproducible set of variables while minimizing errors introduced by LLM-generated recommendations.

## 3.4. Heterogeneous Causal Mediation Analysis

After the treatment variable, outcome, mediator, and adjustment covariates have been identified, the preprocessed dataset and the selected variables are passed to the proposed heterogeneous causal mediation analysis framework, the M-survival learner [Li et al., 2026]. The M-survival learner first estimates the natural indirect treatment effect conditional on covariates (NIECC) for each individual and subsequently constructs a pairwise distance matrix based on the estimated individual-level indirect treatment effects. The distance matrix is then projected into a two-dimensional latent space using tdistributed stochastic neighbor embedding (t-SNE), which preserves the local similarity structure among individuals. Clustering algorithms, such as K-means, are subsequently applied to the low-dimensional representation to identify patient subgroups exhibiting heterogeneous mediation effects. To facilitate clinical interpretation, a decision tree is fitted to the resulting cluster assignments, yielding an explicit and interpretable characterization of the identified subgroup structure. An overview of the heterogeneous causal mediation analysis workflow is presented in Figure 2.

To reduce the risk of identifying spurious subgroup heterogeneity, the framework further incorporates a calibration procedure that distinguishes genuine heterogeneous mediation effects from non-heterogeneous scenarios. Specifically, the calibration step evaluates whether the observed subgroup structure is likely to arise from random variation and filters out subgroup discoveries that do not satisfy the calibration criterion. This procedure improves the reliability and robustness of subgroup identification, thereby reducing the likelihood of misleading clinical conclusions.

### 3.5. Results and Interpretation

To facilitate clinical interpretation, the proposed framework automatically generates a collection of visualization and reporting outputs following the heterogeneous causal mediation analysis. First, an interpretable subgroup profile is produced based on the fitted decision tree, providing explicit decision rules that characterize each identified subgroup using clinically relevant covariates. This representation enables users to understand the defining characteristics of patient subpopulations exhibiting distinct mediation patterns.

To further summarize the estimated heterogeneous mediation effects, boxplots of the estimated NIECC are generated for each identified subgroup. These visualizations illustrate the distribution of individuallevel mediation effects within and across subgroups, allowing users to assess both the magnitude and variability of the indirect treatment effect. Together, the subgroup profiles and NIECC visualizations provide complementary perspectives for interpreting heterogeneous mediation patterns.

To improve the reliability of the reported findings, the framework returns subgroup results only when the calibration procedure indicates statistically credible heterogeneity. If no evidence of heterogeneous mediation effects is detected, no subgroup analysis is reported, thereby reducing the risk of presenting spurious subgroup discoveries. In addition to the visual outputs, the framework generates an automated textual summary describing the identified subgroups, the selected mediator and covariates, and the corresponding clinical interpretation, providing users with a comprehensive and reproducible analysis report. Figure 4 (f) and (g) show the visulization examples.

To facilitate result dissemination and improve analysis reproducibility, SurroPilot automatically generates a comprehensive analysis report upon completion of the heterogeneous causal mediation analysis. The report is organized into four sections: Background, Methods, Results, and Interpretation, providing users with a standardized and reproducible summary of the entire analytical workflow.

The Background section introduces the principles of causal mediation analysis and treatment effect heterogeneity, summarizes the clinical motivation for the analysis, and provides an overview of the uploaded dataset, including variable descriptions and descriptive statistics. This section establishes the analytical context and documents the data used throughout the study.

The Methods section summarizes all user-specified analysis settings, including the treatment and outcome variables, selected mediator, adjustment covariates, and additional domain-specific information provided by the user. It also presents the causal mediation diagram, documents the rationale for selecting the mediator and covariates generated by the LLM, and briefly describes the underlying heterogeneous causal mediation framework.

The Results section reports the identified patient subgroups together with their corresponding decision-tree profiles, subgroup-specific summary statistics, and the estimated NIECC distributions. Figures are automatically generated to visualize the subgroup structure and quantify heterogeneous mediation effects. When no statistically credible heterogeneity is detected through the calibration procedure, the report explicitly states this finding rather than presenting potentially spurious subgroup results.

The Interpretation section provides an integrated clinical interpretation of the analytical results. Specifically, it discusses the biological plausibility of the identified heterogeneous mediation patterns, summarizes their potential clinical implications, highlights important limitations of the analysis, and concludes with recommendations for future investigation. By combining statistical evidence with domainspecific interpretation, the report assists users in translating analytical findings into clinically meaningful insights while maintaining appropriate caution regarding the limitations of observational interpretation.

Overall, the AI-generated report provides a transparent, reproducible, and user-friendly summary of the complete analysis pipeline. By integrating methodological details, quantitative results, visualizations, and automated interpretation into a single document, it reduces the effort required for result reporting and facilitates subsequent clinical research and decision-making. Figure 4 show the example of report. Appendix Figure 5 and 6 illustrate the system prompts used to generate analysis reports when heterogeneous mediation effects are detected and when no significant heterogeneity is detected, respectively. The detailed report example could be found in Appendix C.

## 4. Experiments

SurroPilot is implemented in R for data processing, statistical computation, and causal mediation analysis. To demonstrate its practical utility and evaluate its performance in a real-world clinical trial setting, we applied SurroPilot to the ACTG175 dataset, a landmark Phase III randomized clinical trial that demonstrated the superiority of combination therapy with zidovudine and didanosine over monotherapy with either zidovudine or didanosine in patients with human immunodeficiency virus (HIV) infection [Hammer et al., 1996].

The ACTG175 trial comprised four treatment arms: zidovudine (arm 0), zidovudine plus didanosine (arm 1), zidovudine plus zalcitabine (arm 2), and didanosine (arm 3). In addition to the treatment assignment, survival outcome, and follow-up time, the dataset contains 24 clinical variables that may serve as candidate mediators or adjustment covariates. Identifying appropriate mediators and covariates is often a challenging and time-consuming task for clinical researchers because it requires both statistical expertise and domain knowledge. SurroPilot addresses this challenge by automatically interpreting the dataset and recommending variables for downstream heterogeneous causal mediation analysis. A detailed description of the ACTG175 dataset is provided in Appendix B.

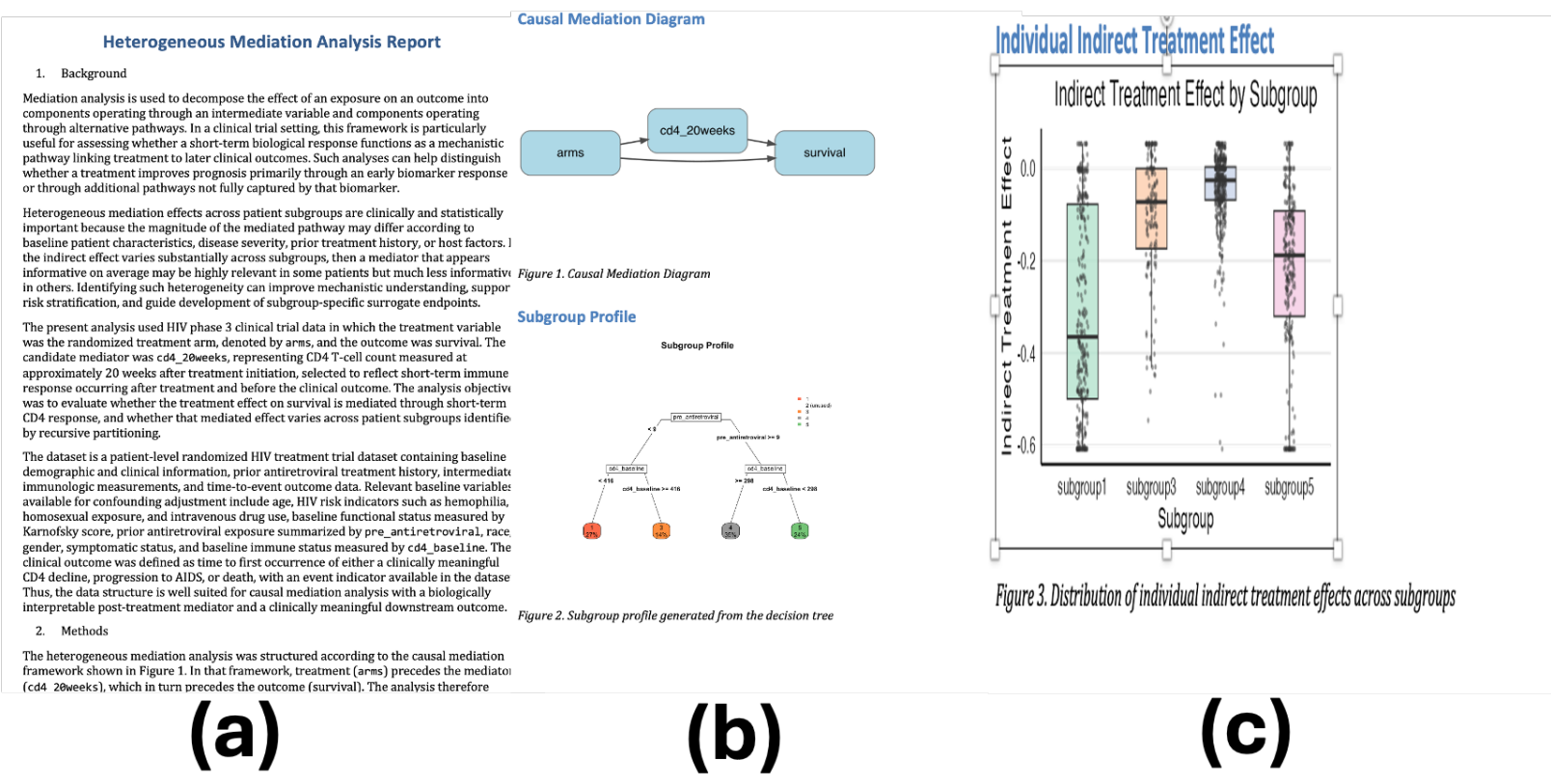

Figure 4 Example of a heterogeneous mediation analysis report generated by SurroPilot. (a) The first page of the textual report. (b) and (c) Representative visualizations generated as part of the report. A complete report example is available in Appendix C.

## 4.1. Comparison of LLMs for Mediator and Covariate Selection

To evaluate the reliability of SurroPilot in identifying mediators and covariates, we compared the variable selections produced by different LLMs with a reference set defined by human experts based on prior clinical knowledge and published literature. Selection performance was evaluated using Precision, Recall, F1 score, and Jaccard similarity. Since many clinical datasets cannot be uploaded to cloud-based services due to privacy regulations, we compared both open-source offline models (Llama-3 and Qwen2.5-14B) and commercial cloud-based models (GPT-5.4-nano and GPT-5.4).

The expert-defined mediator was CD4 count at 20 weeks. The reference covariates set consisted of CD4 baseline, age, weight, hemophilia, homosexual status, Karnofsky score, pre-antiretroviral treatment, race, gender, and symptomatic status. To ensure a fair comparison, all LLMs received the same dataset description, metadata, and task instructions. The complete input prompt is provided in Appendix B.

Jaccard similarity was used to quantify the agreement between the variables selected by an LLM and the expert-defined reference set,

$$\text{Jaccard Similarity} = \frac{|S_{\text{LLM}} \cap S_{\text{Ref}}|}{|S_{\text{LLM}} \cup S_{\text{Ref}}|},$$

where $S_{\text{LLM}}$ and $S_{\text{Ref}}$ denote the variable sets selected by the LLM and by the human experts, respectively. A larger Jaccard similarity indicates greater agreement.

Table 1 shows that larger and more capable language models consistently achieved better agreement with expert selections. For mediator identification, GPT-5.4-nano and GPT-5.4 both achieved perfect performance (Precision = Recall = F1 = Jaccard = 1.00), whereas the open-source models exhibited lower accuracy, particularly Llama-3. This result indicates that advanced LLMs can reliably identify the clinically relevant surrogate endpoint from natural language descriptions.

For covariate selection, GPT-5.4-nano and GPT-5.4 achieved substantially higher recall (0.91 and 0.88, respectively) than the open-source models, indicating that they successfully recovered nearly all expert-selected covariates. Although their precision was moderate (approximately 0.58), suggesting that additional clinically plausible variables were occasionally included, both GPT models achieved the highest overall F1 scores and Jaccard similarities, demonstrating the closest agreement with human experts.

Table 2 further evaluates the identification of two clinically essential covariates, CD4 baseline and preantiretroviral treatment, which are known to be critical for causal adjustment in this dataset. GPT-5.4 correctly identified these variables in over 90% of runs (0.92 for both variables), while GPT-5.4-nano also achieved high recall (0.83 and 0.95, respectively). In contrast, the open-source models showed substantially lower recall, particularly for pre-antiretroviral treatment. These findings suggest that stateof-the-art LLMs are capable of capturing clinically important covaraites with reliability comparable to human experts.

Overall, these results demonstrate that modern LLMs can effectively translate domain knowledge into statistically appropriate variable selections. While commercial models provide the highest accuracy, open-source models offer a practical offline alternative for privacy-sensitive clinical environments where cloud-based services are not permitted.

Table 1 Comparison of mediator and covariate selection performance across different large language models (LLMs). Variable selections were compared with the expert-defined reference using Precision, Recall, F1 score, and Jaccard similarity. Higher values indicate better agreement with human experts.

| | Mediator | | | | Covariate | | | |
|---|---|---|---|---|---|---|---|---|
| Model | Precision | Recall | F1 | Jaccard | Precision | Recall | F1 | Jaccard |
| Llama-3 | 0.63 | 0.76 | 0.67 | 0.63 | 0.40 | 0.13 | 0.19 | 0.12 |
| qwen2.5:14b | 0.97 | 0.97 | 0.97 | 0.97 | 0.76 | 0.33 | 0.44 | 0.30 |
| GPT5.4-nano | 1 | 1 | 1 | 1 | 0.58 | 0.91 | 0.68 | 0.53 |
| GPT5.4 | 1 | 1 | 1 | 1 | 0.57 | 0.88 | 0.69 | 0.53 |

Table 2 Selection frequency of two clinically essential covariates across 100 repeated runs for different LLMs. The values in the second and third columns represent the proportion of runs in which each individual covariate was selected. The final column reports the recall, defined as the proportion of runs in which both essential covariates (CD4 baseline and pre-antiretroviral treatment) were simultaneously identified. Higher values indicate more reliable recovery of the key covariates.

| Model | CD4 baseline Frequency | Pre-antiretroviral Frequency | Joint Recall |
|---|---|---|---|
| Llama-3 | 0.15 | 0.07 | 0.11 |
| qwen2.5:14b | 0.9 | 0.27 | 0.59 |
| Gpt5.4-nano | 0.83 | 0.95 | 0.89 |
| Gpt5.4 | 0.92 | 0.92 | 0.92 |

## 4.2. Comparison of LLMs for Heterogeneous Region Identification

While accurate mediator and covariate selection is important, the ultimate objective of SurroPilot is to identify clinically meaningful heterogeneous patient subgroups. Therefore, we further evaluated whether the heterogeneous regions identified using LLM-selected variables were consistent with those obtained using expert-defined variables after applying the M-Survival Learner.

Figure 5 compares the threshold values of the two clinically essential covariates, pre-antiretroviral treatment and CD4 baseline, across different LLMs. The red triangle denotes the heterogeneous region identified using the expert-defined variables, whereas each gray point corresponds to the thresholds estimated from one independent

LLM-assisted analysis. A gray point is displayed only when both essential covariates were successfully selected by the LLM.

As shown in Figure 5, GPT-5.4 produced threshold values that were highly concentrated around the expert-defined region, demonstrating remarkable consistency across repeated analyses. GPT-5.4-nano also recovered similar heterogeneous regions, although with greater variability and fewer successful runs. In contrast, the open-source models exhibited substantially poorer agreement. Qwen2.5-14B identified the expert-defined region only occasionally, while Llama-3 rarely selected the essential covariates simultaneously, resulting in almost no successful recovery of the heterogeneous region.

To further quantify this agreement, Table 3 reports the number of successful recoveries of the expertdefined heterogeneous region over 100 independent runs. GPT-5.4 achieved agreement in 92 out of 100 runs, substantially outperforming GPT-5.4-nano (42 runs), Qwen2.5-14B (23 runs), and Llama-3 (1 run). These findings indicate that advanced LLMs can reliably translate clinical knowledge into statistically appropriate variable selections and subsequently reproduce heterogeneous subgroup analyses that closely match those obtained by domain experts.

Overall, these results highlight the effectiveness of the human–AI collaborative framework adopted by SurroPilot. Domain experts provide clinical knowledge when available, while the LLM assists with data understanding, variable selection, and workflow automation. Combined with the validated M-Survival Learner, this collaborative design enables reliable heterogeneous subgroup identification that closely matches expert analyses, thereby making advanced causal mediation analysis more accessible to clinical researchers.

Besides, we also evaluate the inspector agent which can be found in Appendix B.

Table 3 Comparison of heterogeneous covariates selection performance across different LLMs.

| Model | frequency with human's results |
|---|---|
| Llama-3 | 1 |
| qwen2.5:14b | 23 |
| gpt5.4-nano | 42 |
| gpt5.4 | 92 |

## 5. Example

We demonstrate the use of SurroPilot for identifying heterogeneous surrogate biomarkers using the ACTG175 HIV Phase III clinical trial dataset. The complete analysis workflow is illustrated in Figure 6, including data understanding and preparation, variable identification, and heterogeneous causal mediation analysis. The inputs are the same with the Section 4. To further facilitate reproducibility and practical adoption, we also provide a video demonstration that showcases the use of SurroPilot for clinical trial data analysis (https://www.youtube.com/watch?v=hh9eMdok2mE).

## 6. Discussion

SurroPilot extends the role of large language models in clinical trials from operational assistance to AIassisted statistical inference. By integrating natural-language interaction with validated heterogeneous causal mediation methodology, the proposed framework enables end-to-end surrogate endpoint evaluation while preserving statistical rigor through a human-in-the-loop design. Rather than replacing statistical methodology or domain expertise, SurroPilot combines LLM-assisted reasoning with validated causal inference procedures to improve the accessibility, reproducibility, and efficiency of heterogeneous surrogate biomarker analysis. Experimental results from the ACTG175 Phase III clinical trial further demonstrate that modern LLMs can reliably assist with clinically relevant analytical tasks, supporting the feasibility of integrating foundation models into advanced statistical workflows.

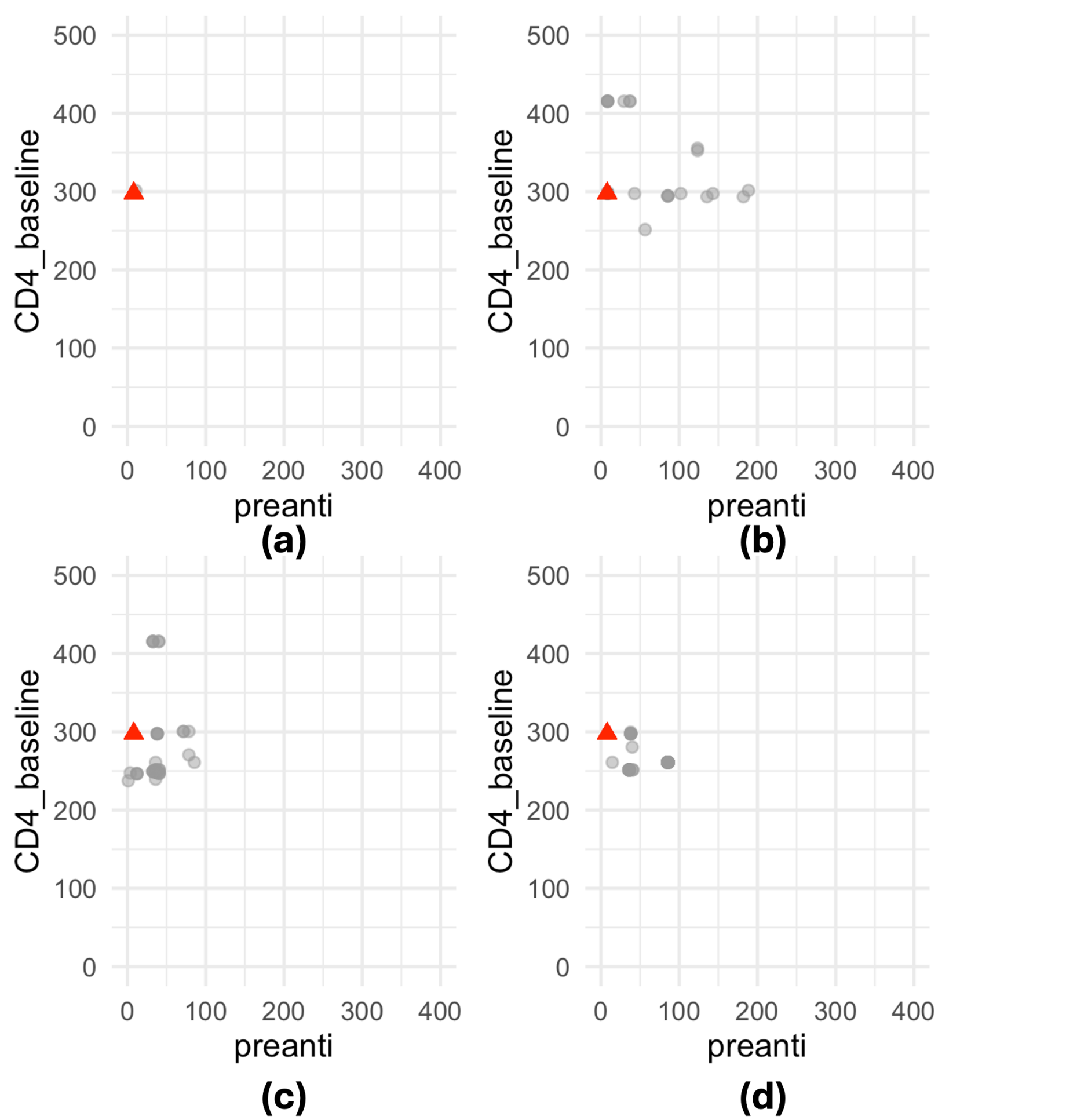

Figure 5 Comparison of heterogeneous regions identified using variables selected by different LLMs. Panels (a)–(d) correspond to Llama-3, Qwen2.5-14B, GPT-5.4-nano, and GPT-5.4, respectively. The red triangle represents the threshold values of the expert-defined heterogeneous region based on the two clinically essential covariates, CD4 baseline and pre-antiretroviral treatment (preanti). Each gray point corresponds to the threshold values obtained from one of 100 independent LLM-assisted analyses and is shown only when both essential covariates were selected. Greater concentration of gray points around the red triangle indicates stronger agreement with the expert-defined heterogeneous region.

A key contribution of this work is the integration of AI-assisted statistical computing with heterogeneous causal mediation analysis. Existing LLM-assisted applications in clinical trials have primarily focused on operational tasks, such as protocol development, patient recruitment, data extraction, and clinical documentation, whereas statistical inference has remained largely beyond the scope of AI-assisted systems. SurroPilot demonstrates that LLMs can effectively assist with data understanding, preprocessing, mediator and covariate selection, statistical workflow configuration, and result interpretation, while the underlying causal estimation is performed using validated statistical methodology. This collaborative paradigm substantially lowers the technical barriers to heterogeneous mediation analysis without compromising methodological transparency, reproducibility, or expert oversight.

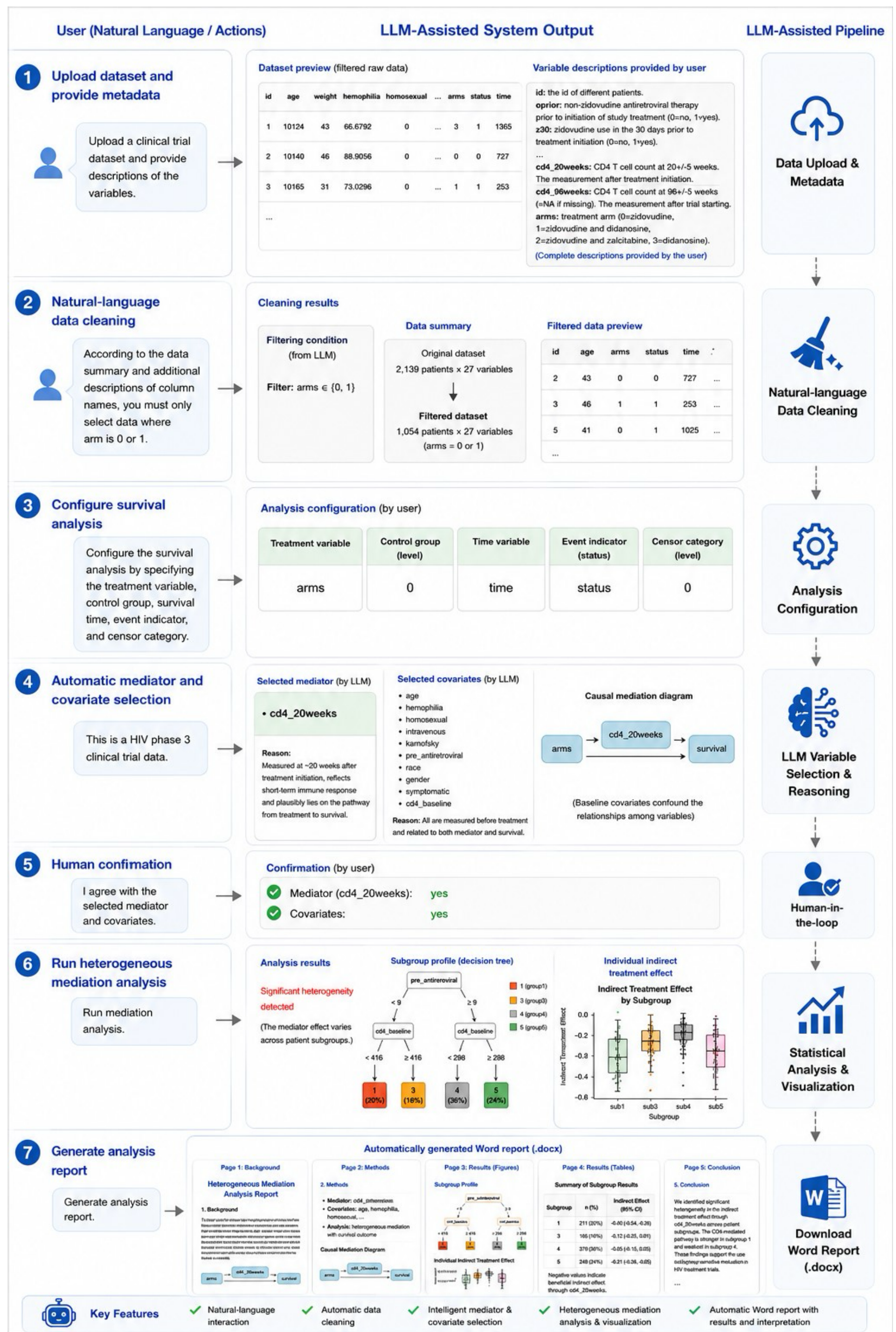

Figure 6 Example of using the proposed LLM-assisted heterogeneous mediation analysis system ”SurroPilot”. Users interact with the system entirely through natural-language instructions. The system supports dataset upload, metadata understanding, user-directed data cleaning, survival analysis configuration, automatic mediator and covariate selection, human confirmation, heterogeneous mediation analysis, visualization of subgroup-specific effects, and automatic generation of a downloadable Word report containing the statistical methods, results, figures, and interpretation.

Beyond improving analytical efficiency, SurroPilot enables subgroup-specific surrogate endpoint evaluation, which has important implications for precision medicine and clinical trial design. Conventional surrogate validation relies primarily on population-average measures and implicitly assumes homogeneous surrogate validity across patients. However, biological mechanisms, treatment responses, and disease progression frequently vary across clinically meaningful patient subgroups. By integrating heterogeneous mediation analysis with interpretable subgroup identification, SurroPilot enables investigators to identify patient populations in which a surrogate biomarker provides stronger or weaker evidence for long-term clinical benefit. Rather than asking whether a biomarker is a valid surrogate on average, the proposed framework addresses a clinically more informative question: for whom is the surrogate valid? Such subgroup-specific evaluation has the potential to improve patient stratification, biomarker-driven clinical trial design, regulatory decision-making, and precision medicine.

Another important finding is that state-of-the-art LLMs achieved high agreement with expertdefined mediator and covariate selections. These results suggest that LLMs are capable of translating domain knowledge expressed in natural language into statistically appropriate analytical decisions. Importantly, SurroPilot does not rely on autonomous decision-making. Instead, all recommendations remain transparent, explainable, and subject to user confirmation before downstream analyses are performed. This human–AI collaborative paradigm provides a practical approach for integrating foundation models into rigorous statistical workflows while maintaining scientific accountability.

Several limitations should be acknowledged. First, the current implementation is primarily designed for randomized Phase III clinical trials and therefore has not been extensively evaluated in observational studies or other clinical research settings. Second, although the proposed programmer–inspector framework substantially improves the reliability of AI-generated code and variable selection, LLMs may still occasionally generate inaccurate recommendations or unsupported analyses, making human verification essential throughout the analytical process. Third, the current framework focuses on structured clinical datasets and has not yet incorporated multimodal information, such as medical imaging, genomics, electronic health records, or external biomedical knowledge bases. Finally, the statistical workflow currently centers on heterogeneous causal mediation analysis, and further investigation is needed to determine whether similar AI-assisted statistical computing paradigms can be generalized to other classes of causal inference and statistical learning methods.

Future work will focus on extending both the statistical methodology and the AI capabilities of SurroPilot. From a methodological perspective, incorporating Bayesian mediation models, historical borrowing methods, adaptive clinical trial designs, and real-world evidence may further improve the precision and efficiency of heterogeneous surrogate evaluation, particularly for rare patient subgroups [Hobbs et al., 2011, Zhou et al., 2026]. From an AI perspective, integrating more advanced reasoning models, domain-specific biomedical foundation models, and multimodal LLMs may further enhance the robustness, reliability, and interpretability of AI-assisted statistical computing. Continued improvements in user interaction, visualization, and report generation will further facilitate the adoption of advanced statistical methodologies by clinicians and biomedical researchers.

In conclusion, SurroPilot demonstrates the feasibility of integrating large language models with validated statistical methodology for heterogeneous surrogate endpoint evaluation. By combining AIassisted statistical computing with human expertise, the proposed framework substantially lowers the barriers to advanced causal mediation analysis while maintaining interpretability, reproducibility, transparency, and statistical rigor. More broadly, this work illustrates that large language models can serve as intelligent interfaces for complex statistical inference rather than replacements for statistical methodology, providing a promising direction for the next generation of AI-assisted statistical analysis tools in biomedical research.

## Code Availability

The R Shiny web application will be available online soon. The source code is available to reviewers via an anonymous OSF view-only link: https://osf.io/turb4/?view_only= abfa2480479d4282a216ba11c73d6ff0. A demonstration video is available at https://www.youtube.com/ watch?v=hh9eMdok2mE.

## Author Contributions Statement

X.L. and P.W. contributed to conception and design of the study. X.L. organized the database and performed the statistical analysis. X.L. and P.W. wrote the first draft of the manuscript. All authors contributed to manuscript revision, read, and approved the submitted version.

## CONFLICT OF INTEREST

None.

## Acknowledgments

P.W. was partially supported by the National Institutes of Health grants R21HL170213, R01HL18406 and P01CA296429. The funders play no role in the study design, data collection and analysis, decision to publish, or preparation of the manuscript.